\documentclass[a4paper]{jpconf}
\usepackage{graphicx}
\usepackage{lineno}
\usepackage{algorithmic}
\usepackage{hyperref}
\usepackage{xcolor}
\usepackage{afterpage}
\usepackage{subcaption}
\usepackage[T1]{fontenc}

\usepackage[inline]{enumitem}   
\makeatletter
\newcommand{\inlineitem}[1][]{%
\ifnum\enit@type=\tw@
    {\descriptionlabel{#1}}
  \hspace{\labelsep}%
\else
  \ifnum\enit@type=\z@
       \refstepcounter{\@listctr}\fi
    \quad\@itemlabel\hspace{\labelsep}%
\fi}

\definecolor{id}{HTML}{A6CE39}

\newenvironment{shadequote}{%
  \begin{center}%
    \begin{minipage}{.8\linewidth}%
      \begin{shaded}%
        \sffamily\slshape}{%
      \end{shaded}
    \end{minipage}%
  \end{center}%
}

\usepackage{color}
\usepackage{framed}
\definecolor{shadecolor}{gray}{0.95}

\hypersetup{
colorlinks = true
}

\newcommand{\AtlasCoordFootnote}{%
ATLAS uses a right-handed coordinate system with its origin at the nominal interaction point in the centre of the detector and the $z$-axis along the beam pipe.
The $x$-axis points from the interaction point to the centre of the LHC ring,
and the $y$-axis points upwards.
Cylindrical coordinates $(r,\phi)$ are used in the transverse plane, 
$\phi$ being the azimuthal angle around the $z$-axis.
The pseudorapidity is defined in terms of the polar angle $\theta$ as $\eta = -\ln \tan(\theta/2)$.
}

\begin{document}
\title{Machine Learning Algorithms for $b$-Jet Tagging at the ATLAS Experiment}

\author{Michela Paganini 
\\on behalf of the ATLAS Collaboration}

\address{Yale University, Department of Physics, 217 Prospect Street, New Haven, CT 06511-8499}

\ead{\href{mailto:michela.paganini@cern.ch}{michela.paganini@cern.ch}}

\begin{abstract}
The separation of $b$-quark initiated jets from those coming from lighter quark flavors ($b$-tagging) is a fundamental tool for the ATLAS physics program at the CERN Large Hadron Collider. The most powerful $b$-tagging algorithms combine information from low-level taggers, exploiting reconstructed track and vertex information, into machine learning classifiers. The potential of modern deep learning techniques is explored using simulated events, and compared to that achievable from more traditional classifiers such as boosted decision trees.
\end{abstract}

\section{Introduction}

\begin{figure}[b]
    \centering
    \includegraphics[width=0.4\textwidth]{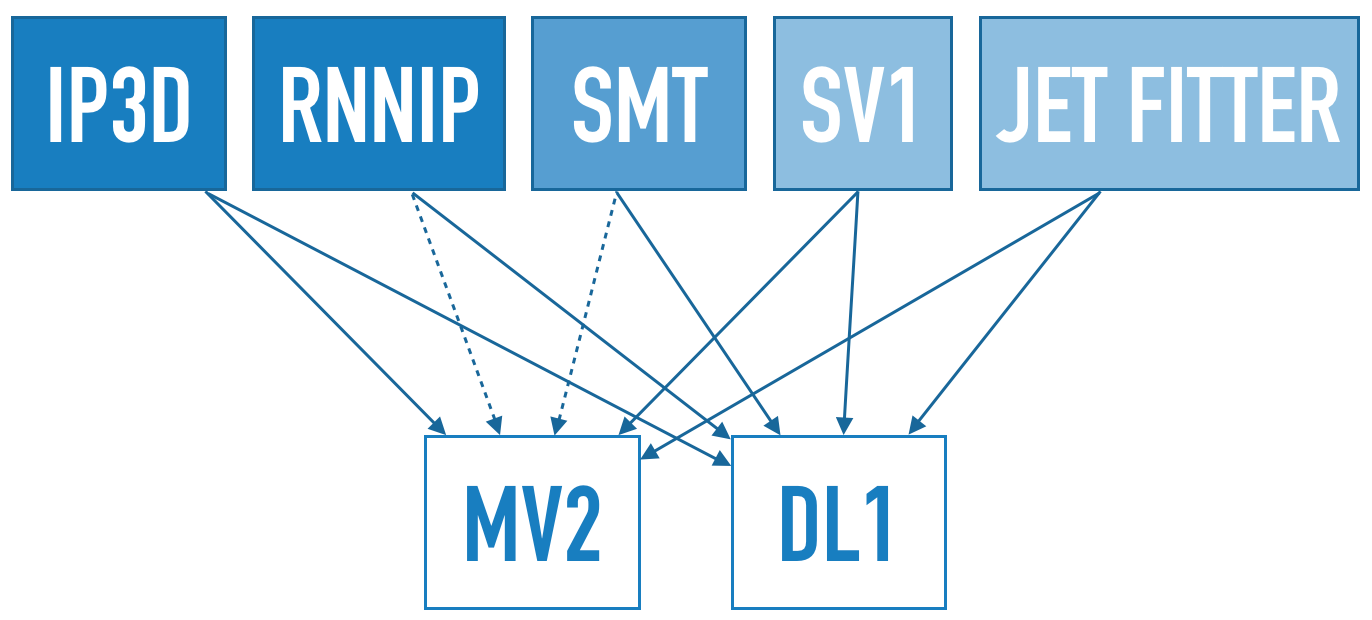}
    \caption{The set of $b$-tagging algorithms currently in use in the ATLAS experiment at the LHC includes five low-level taggers (\textsc{IP3D}, \textsc{RNNIP}, \textsc{SMT}, \textsc{SV1}, and \textsc{JetFitter}), as well as two high level taggers (\textsc{MV2}, and \textsc{DL1}) which take as inputs combinations of the outputs of the low-level taggers.}
    \label{fig:eco}
\end{figure}

With the increase in the expected integrated luminosity delivered in the remaining years of Run 2 by the Large Hadron Collider~\cite{lhc} (LHC), a growing emphasis is put on algorithmic improvements at the reconstruction level~\cite{recorun2} to leverage the full sensitivity potential for measurements and searches of interest to the ATLAS~\cite{ATLAS} physics program. 

The task of identifying jets that contain $b$-hadrons and separating them from jets initiated by lighter quark flavors is of particular significance due to the prominence of $b$-hadrons in final states of interesting physics processes, and to the abundant background light-flavored-jet production at hadron colliders such as the LHC. This task, commonly referred to as $b$-tagging, is broadly applied to both precise standard model measurements and searches for signatures of new physics phenomena. $b$-tagging can be cast as a classification problem with the objective of correctly assigning jet flavors using the trajectories of charged particles (tracks) reconstructed in the inner detector~\cite{atlasID}, and then associated to jets reconstructed from clusters of energy in the electromagnetic~\cite{larcalo} and hadronic~\cite{tilecalo} calorimeters. 

We present the current hierarchy of algorithms used for $b$-tagging in ATLAS (Figure~\ref{fig:eco}), and outline the latest Monte Carlo-based optimization results~\cite{ATL-PHYS-PUB-2017-013}, with specific focus on deep learning-powered improvements, which include the use of fully-connected and recurrent neural networks. These are the algorithms to be used to analyze the data recorded in 2017-2018.


\section{Heavy hadron topology}
\label{heavyhadrontopo}
The typical heavy hadron topology presents one or more vertices that are displaced from the hard-scatter interaction point. 
The longer lifetimes of heavy-flavor hadrons correspond to macroscopic decay lengths that can be resolved by the ATLAS vertexing detectors. The hadron's decay is governed by its heavy quark decay chain. A $b$-hadron will usually decay through a cascade to a $c$-hadron, strange hadron, etc. Therefore, in a $b$-initiated event, the goal is to reconstruct not only the primary vertex whence the event initiates, but also a secondary, and possibly tertiary vertex along the decay cascade. Finally, the heavy-flavor hadrons' non-negligible branching ratio to semi-leptonic decays ($\approx 21\%$) suggests using the presence of soft muons within jets as yet another discriminating feature for flavor tagging.

Jets used for $b$-tagging are reconstructed using the anti-$k_t$ algorithm~\cite{akt} with radius $R=0.4$, and pass the following selection criteria~\cite{ATL-PHYS-PUB-2017-013}:
\begin{itemize}
    \item $p_\mathrm{T}^\mathrm{jet} > 20$ GeV
    \inlineitem $\left| \eta^\mathrm{jet} \right| > 2.5$
    \inlineitem if $\left| \eta^\mathrm{jet} \right| < 2.4$ and $p_\mathrm{T}^\mathrm{jet} < 60$ GeV: JVT$^\mathrm{jet}$ > 0.59
\end{itemize}
where JVT is the output of the Jet Vertex Tagger algorithm~\cite{jvt} used to suppress jets from pile-up interactions. At training time, the selection cuts are applied to uncalibrated jets, while at test time the kinematic inputs are calibrated.

Tracks are associated to the nearest jet based on an angular separation criterion that varies as a function of the jet $p_\mathrm{T}$~\cite{btagperf}.

At truth level, using the kinematic properties of the jet under investigation and the hadrons in the event, the logic of the devised labeling scheme goes as follows:
\begin{shadequote}
\begin{algorithmic}
\IF{$(\Delta R(\mathrm{jet}, \ B) < 0.3)\  \& \  (p_{T_B} > 5 \ \mathrm{GeV})$} 
\STATE {jet flavor $\gets b$}
\ELSIF{$(\Delta R(\mathrm{jet}, \ C) < 0.3)\  \& \  (p_{T_C} > 5 \ \mathrm{GeV})$}
\STATE{jet flavor $\gets c$}
\ELSIF{$(\Delta R(\mathrm{jet}, \ \tau) < 0.3)\  \& \  (p_{T_\tau} > 5 \ \mathrm{GeV})$} \STATE{jet flavor $\gets \tau$}
\ELSE 
\STATE{jet flavor $\gets$ light} 
\ENDIF
\end{algorithmic}
\end{shadequote}
where $\Delta R (a,b) \equiv \sqrt{(\phi_a - \phi_b)^2 + (\eta_a - \eta_b)^2}$ \footnote{\AtlasCoordFootnote} and $p_\mathrm{T}$ identifies the component of the momentum that is perpendicular to the beam direction.

\section{Low-level taggers}
Given the unique topology of heavy hadron decays (Section~\ref{heavyhadrontopo}), ATLAS deploys a suite of low-level $b$-tagging algorithms that aim at exploiting different discriminative aspects of $b$-quark jets.
\subsection{Secondary vertex-based $b$-tagging algorithms}
Physically grounded algorithms have been designed to look for displaced vertices. ATLAS uses an inclusive secondary vertex algorithm, \textsc{SV1}~\cite{ATL-PHYS-PUB-2017-011}, which looks for displaced tracks and groups them in a secondary vertex, and a multi-vertex finding algorithm, \textsc{JetFitter}~\cite{jf}, which reconstructs the topology and fits the entire decay chain along the hadron line of flight. 

\subsection{Impact parameter-based $b$-tagging algorithms}
Another low-level $b$-tagging approach exploits the impact parameter properties of tracks and how they differ in the heavy flavor and light flavor scenarios. The goal is to compute the likelihood of tracks associated to jets originating at the primary vertex. 

We define the transverse impact parameter, $d_0$, as the distance of closest approach of the track to the primary vertex in the $r-\phi$ plane (the plane perpendicular to the beam direction $\hat{z}$), and the longitudinal impact parameter, $z_0$, as the distance from the primary vertex to the intersection of the track projection onto the longitudinal plane with the beam axis~\cite{btagging16}. The sign of the impact parameters indicates whether the track crosses the jet axis in front of or behind the primary vertex with respect to the jet direction of flight. While tracks in light jets have impact parameter significances that are close to zero, those generated from $b$-hadron decays and consistent with the secondary vertex hypothesis are more likely to deviate from zero~\cite{btagging16}.

The tracks used for impact parameter-based tagging satisfy:
\begin{itemize}
    \item $p_\mathrm{T}^\mathrm{track} > 1$ GeV
    \item $|d_0| <1$ mm and $|z_0$ sin $\theta| <1.5$ mm
    \item $N_\mathrm{Si}^\mathrm{hits} \geq 7$ and $N_\mathrm{Si}^\mathrm{holes} \leq 2$ with $N_\mathrm{pixel}^\mathrm{holes} \leq 1$
\end{itemize}
where a hole is defined as a missing expected hit and the subscripts on $N$ identify different inner detector components.

\subsubsection{IP3D}

Traditionally, impact parameter tagging uses a binned 2D likelihood method in different track grade categories using the transverse and longitudinal impact parameter significances. The \textsc{IP3D} tagger is based on a log likelihood ratio discriminant, computed -- assuming no correlation among tracks -- as the following sum of per-track contributions:
\begin{equation}
\label{eq:ip3d}
\mathrm{IP3D \ LLR} = \sum_{i=1}^N \mathrm{log}\frac{p_{b_i}}{p_{u_i}}
\end{equation}
where $N$ is the number of tracks associated to the jet, and $p_b$, $p_u$ are the template PDFs for the $b$ and light flavor hypotheses. 

However, in a $b$-hadron decay, several charged particles with large impact parameters can emerge from the secondary (or tertiary) vertex. In this scenario, the tracks’ impact parameters are intrinsically correlated: once a track with large impact parameter is found, finding a second track with large impact parameter becomes more likely. On the other hand, such a correlation should not be observed if there is no displaced decay, which is often the case for light-flavor jets. \textsc{IP3D}'s treatment of tracks as independent limits its power as its formulation (Eq.~\ref{eq:ip3d}) disregards known track correlations.

\subsubsection{RNNIP}
The traditional \textsc{IP3D} tagger has been augmented by the introduction of \textsc{RNNIP}~\cite{rnnip}, a Recurrent Neural Network-based $b$-tagging classifier.

\begin{figure}
\begin{minipage}{0.45\linewidth}
    \centering
    \includegraphics[height=4.5cm]{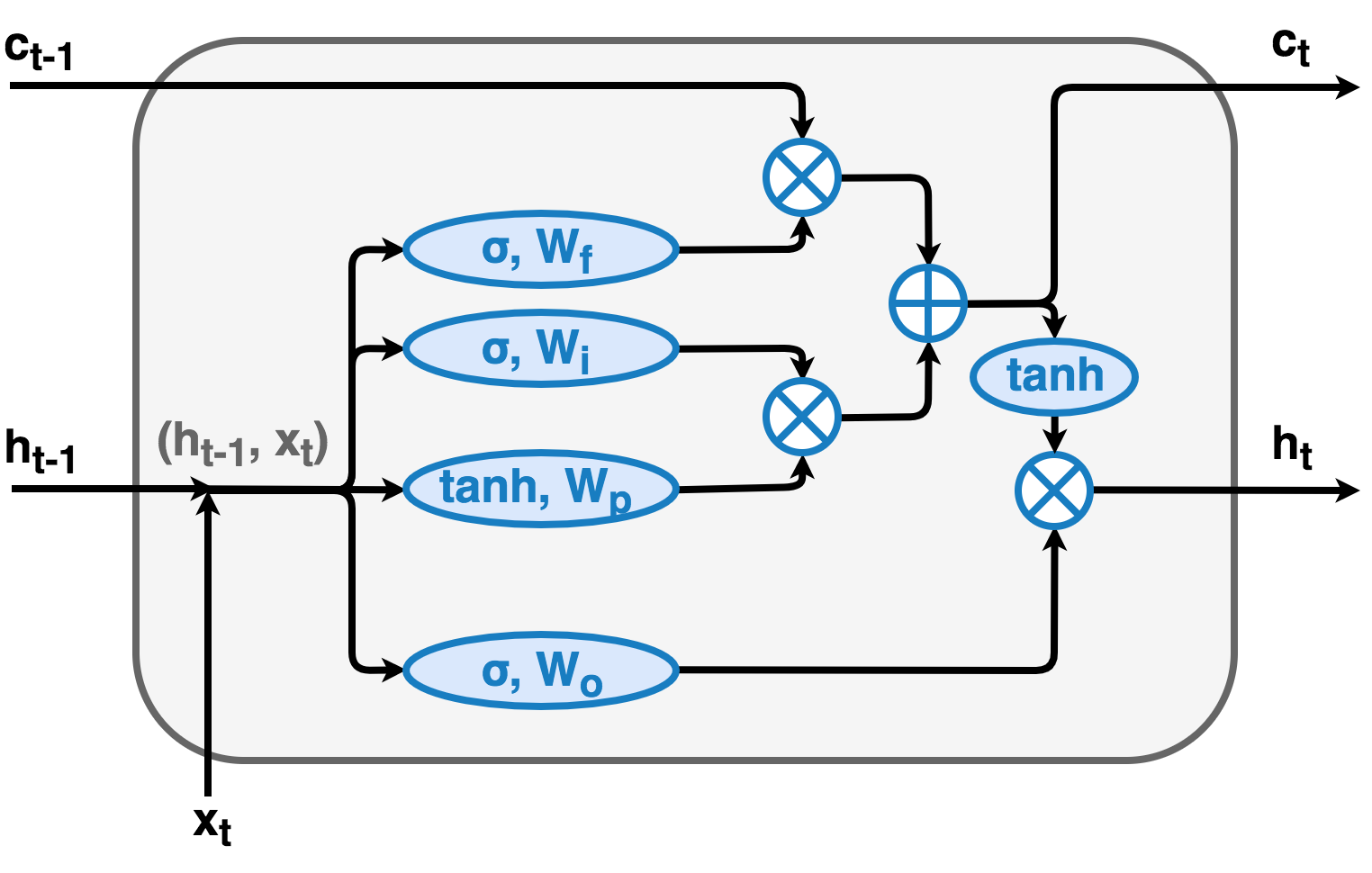}
    \caption{Anatomy of an LSTM cell with output, input, and forget gates.}
    \label{fig:lstm}
\end{minipage}\hfill
\begin{minipage}{0.45\linewidth}
    \centering
    \includegraphics[height=4.5cm]{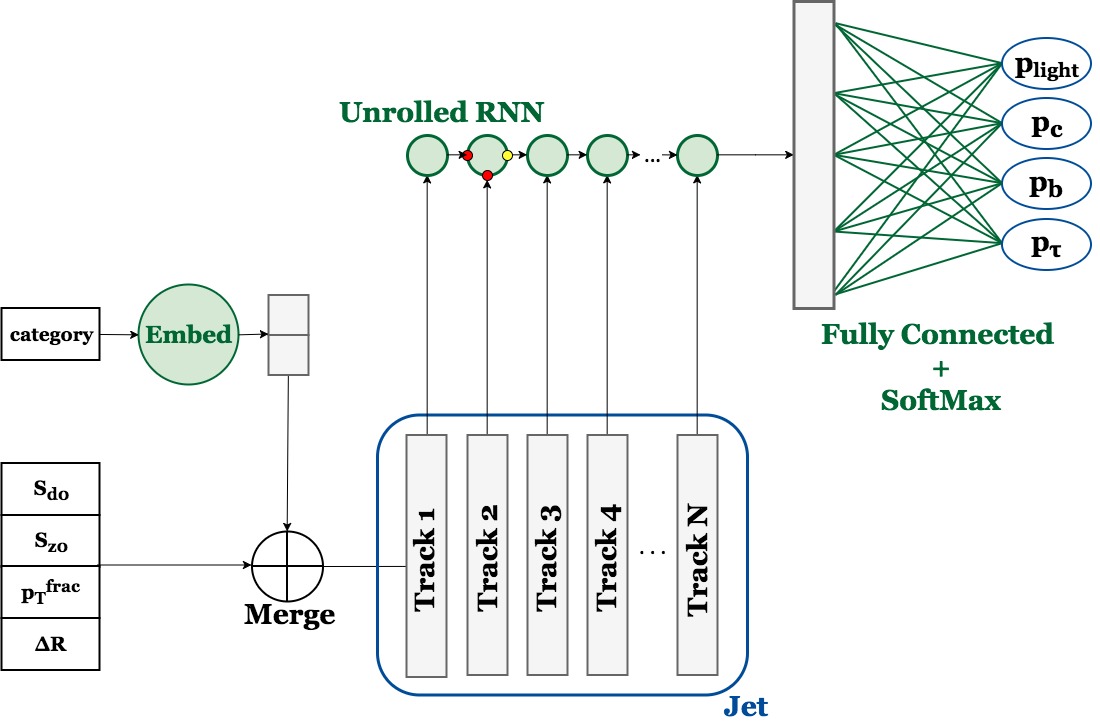}
    \caption{Diagram of the \textsc{RNNIP} tagger from input (left) to output (right).}
    \label{fig:rnnip}
\end{minipage}
\end{figure}

Recurrent Neural Networks (RNNs) are generally used to learn sequence-based dependencies for arbitrary-length input sequences. The fundamental unit is a cell that holds an internal state vector.
To allow information to persist, RNNs use loops that feed the output of the cell back into the cell itself at the following step. Much of the recent success of RNNs comes from the formulation of Long Short-Term Memory~\cite{lstm} (LSTM) units and later variants such as Gated Recurrent Units~\cite{gru} (GRUs). These modifications at the cell level mitigate issues related to vanishing and exploding gradients, to improve the knowledge persistence of long-term dependencies. They do so with different internal gating mechanisms that modify the cell state in order to regulate the relative importance of long-term versus short-term information, allowing the unit to learn when to read, write and reset the memory. The inner structure of an LSTM unit is depicted in Figure~\ref{fig:lstm}, while Figure~\ref{fig:rnnip} shows its usage in the context of the tagger.

This architecture is the natural candidate for processing inputs in sequence format. \textsc{RNNIP} represents jets as a sequence of tracks ordered by their absolute transverse impact parameter significance. Each track is described by a vector of features: transverse and longitudinal impact parameter significances, jet $p_\mathrm{T}$ fraction carried by the track, distance between the track and the jet axis, and a learned 2D embedding of the track quality.

This four-class tagger predicts a jet flavor probability associated with each flavor ($b$, $c$, $\tau$, and light), which could be used to select a class using a maximum a posteriori estimation. 
For the purpose of calibrating multiple efficiency points, this information is instead combined into the discriminant
\begin{equation}
\mathrm{RNNIP}(f_c, f_\tau) = \mathrm{log}\left(\frac{p_b}{f_c \cdot p_c + f_\tau \cdot p_\tau + (1-f_c-f_\tau) \cdot p_u} \right)
\end{equation}
where each background importance can be varied by varying their fraction ($f_c$, $f_\tau$) after training.

\subsection{Muon-based $b$-tagging algorithm}
The Soft Muon Tagger~\cite{ATL-PHYS-PUB-2017-013} (\textsc{SMT}) relies on information from the reconstructed muons from semi-leptonic decays of heavy-flavor hadrons. \textsc{SMT} provides orthogonal information to impact parameter- and vertex-based taggers. This tagger uses a group of 6 hand-engineered variables and combines them in a boosted decision tree to obtain a discriminant.

\section{High-level taggers}
\subsection{MV2}
The recently improved \textsc{MV2} tagger is a gradient boosted decision tree trained using the \texttt{ROOT TMVA} library~\cite{TMVA} to perform a binary classification of jets into $b$ and non-$b$. Three variants are investigated which differ in the quantity of input variables.

\noindent
\textbf{\textsc{MV2}}: this baseline model uses 24 input variables from lower level taggers (\textsc{IP2D}, \textsc{IP3D}, \textsc{SV1}, \textsc{JetFitter}) and kinematic properties. The full list is available in Ref.~\cite{ATL-PHYS-PUB-2015-022};

\noindent
\textbf{\textsc{MV2Mu}}: adds the output of the \textsc{SMT} BDT to the list of inputs;

\noindent
\textbf{\textsc{MV2MuRnn}}: adds the output of \textsc{RNNIP} to the list of inputs.

Importance sampling is performed to achieve a flat 2D distribution in $p_\mathrm{T}$ and $\eta$. The hyper-parameters of the model were optimized for the 2016 Run~\cite{btagging16} to 1000 trees with maximum depth of 30, and minimum node size of 0.05\% of the training sample.

\subsection{DL1}
\textsc{DL1}~\cite{dl1} is deep neural network whose architecture is a mixture of fully-connected, maxout~\cite{maxout}, and batch normalization~\cite{batchnorm} layers with rectified linear units~\cite{relu} as activation functions. Dropout~\cite{dropout} is used as a stochastic regularization technique. 

As with \textsc{RNNIP}, \textsc{DL1} is trained using \texttt{Keras}~\cite{keras}, and incorporated into the ATLAS Athena code base~\cite{athena} via the custom C++ network evaluator provided through the \texttt{LWTNN} library~\cite{lwtnn}. \texttt{rootpy}~\cite{rootpy} and \texttt{root\char`_numpy}~\cite{root_numpy} were used to convert the input file format.

The network is trained to minimize the cross-entropy loss using the Adam optimizer~\cite{adam}. A thorough grid search is performed to select the network structure and hyper-parameters such as the number of hidden layers, number of nodes per layer, and the learning rate.

The three output nodes calculate the probabilities associated with each jet flavor (here: $p_b$, $p_c$, $p_u$) as highly non-linear functions of the input features. A final discriminant for $b$-tagging is obtained by combining the outputs into one tunable function of the fraction of $c$-jets in the background, $f_c$:
\begin{equation}
\mathrm{DL1}(f_c) = \mathrm{log}\left(\frac{p_b}{f_c \cdot p_c + (1-f_c) \cdot p_u} \right)
\end{equation}
A similar equation can be derived for $c$-tagging, by swapping the $b$ and $c$ labels. Although the $c$-fraction can be varied depending on the background composition of each analysis, in order to facilitate the comparison with \textsc{MV2}, $f_c$ is herein set to the natural occurring fraction in a typical $t\bar{t}$ sample, \textit{i.e.} 7\%. This parameter regulates the trade-off between $c$ and light rejections at a given $b$ tagging efficiency.

While \textsc{DL1} and \textsc{MV2} have been found to have similar performance, \textsc{DL1} has emerged out of need for improved flexibility and power in future R$\&$D cycles. The neural network can be easily trained with minimal standalone code; it is GPU-enabled; it is modular and can be effortlessly extended to include more input variables. It is also amenable to be trained using an adversarial loss to minimize data-Monte Carlo discrepancies~\cite{domain-adv}, and it grants the possibility for future end-to-end trainings in conjunction with \textsc{RNNIP}. 

\section{Performance in Simulation}
The high level taggers are trained on a newly introduced hybrid sample which combines $t\bar{t}$ with a broad $Z'$ sample at $m_{Z'} = 4$ TeV to increase the representation of high $p_\mathrm{T}$ jets at training time.

\begin{figure}
\begin{subfigure}{0.45\textwidth}
    \centering
    \includegraphics[width=\textwidth]{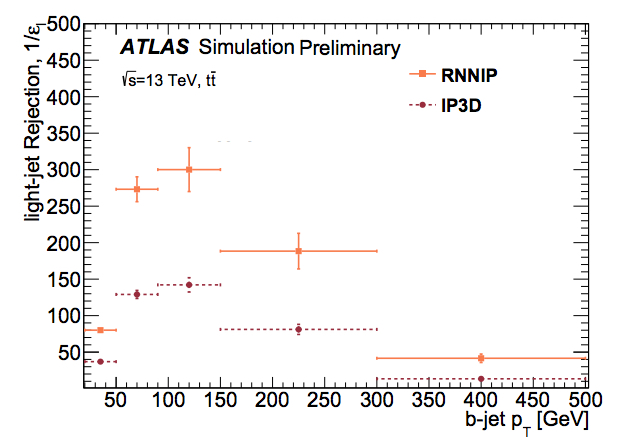}
    \caption{\ }
\end{subfigure}
\hfill
\begin{subfigure}{0.45\textwidth}
    \centering
    \includegraphics[width=\textwidth]{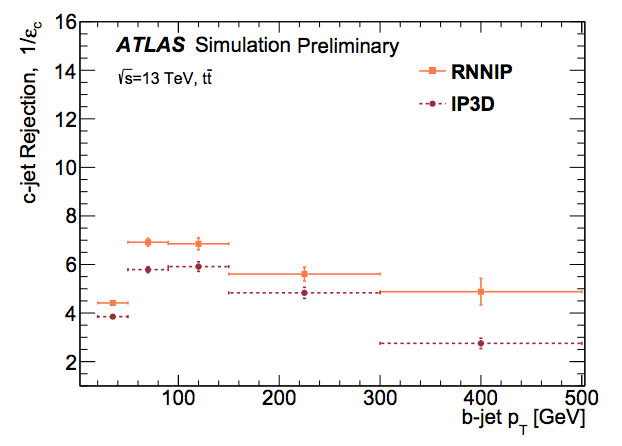}
    \caption{\ }
\end{subfigure}
\caption{Performance comparison in bins of $p_\mathrm{T}$, as it appeared in Ref.~\cite{ATL-PHYS-PUB-2017-013}, between \textsc{IP3D} and \textsc{RNNIP} for light jet (a) and $c$-jet rejection (b) in simulated $t\bar{t}$ events at a constant $b$-efficiency of 70\%.}
\label{fig:rnnvsip3d}
\end{figure}

Figure~\ref{fig:rnnvsip3d} shows that \textsc{RNNIP} improves light rejection by a factor of 2 and $c$ rejection by a factor of 1.2 compared to \textsc{IP3D}, and outperforms it across the entire $p_\mathrm{T}$ spectrum. A sizable fraction of $b$-jets is tagged correctly by only one of the two algorithms, where \textsc{IP3D} is more efficient at low track multiplicity and short decay length, while \textsc{RNNIP} is superior at high multiplicity.

\begin{figure}
\centering
\begin{subfigure}{0.37\textwidth}
    \centering
    \includegraphics[width=\textwidth]{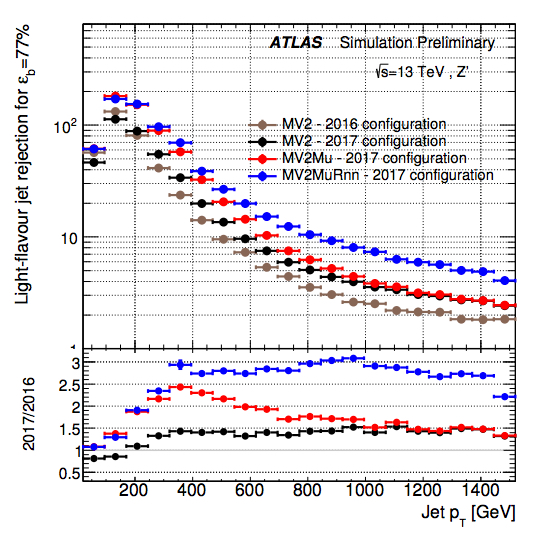}
    \caption{\ }
\end{subfigure}
\hspace{0.07\textwidth}
\begin{subfigure}{0.37\textwidth}
    \centering
    \includegraphics[width=\textwidth]{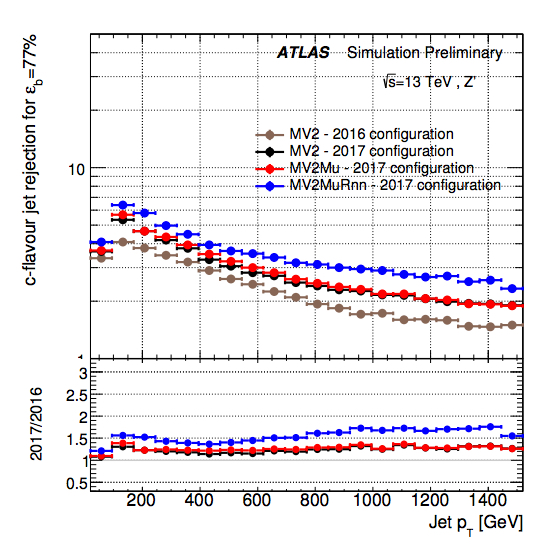}
    \caption{\ }
\end{subfigure}
\caption{Performance comparison in bins of $p_\mathrm{T}$, as it appeared in Ref.~\cite{ATL-PHYS-PUB-2017-013}, between \textsc{MV2} in the 2016 configuration (brown), \textsc{MV2} in the 2017 configuration (black), \textsc{MV2Mu} (red) and \textsc{MV2MuRnn} (blue) for light jet (a) and $c$-jet rejection (b) in simulated $Z'$ events at a constant $b$-efficiency of 77\%. The ratio plots use the 2016 \textsc{MV2} configuration as baseline.}
\label{fig:mv2perf}
\end{figure}

The three \textsc{MV2} versions are evaluated on exclusive $t\bar{t}$ and $Z'$ samples; results from the latter are presented in Figure~\ref{fig:mv2perf}. Including muon information improves $b$-to-light jet separation, but the improvement vanishes at high $p_\mathrm{T}$.
On the other hand, $b$-to-$c$ separation does not strongly benefit from the muon-in-jet information for efficiencies > 50\% because of lepton presence in both $b$ and $c$ decays. \textsc{MV2MuRnn} further improves performance, especially in the high-$p_\mathrm{T}$ region where the \textsc{RNNIP} tagger is most performant. For example, when evaluating on $Z'$ events (Figure~\ref{fig:mv2perf}), at $p_\mathrm{T}^\mathrm{jet} = 1$ TeV the RNN improves the light-jet rejection by a factor $\sim 1.5$ and the $c$-jet rejection by a factor of $\sim 1.3$, compared to the baseline \textsc{MV2} 2017 configuration.

\section{Conclusion}
ATLAS developed new algorithms to identify $b$-jets using machine learning techniques. Their performance was investigated using Monte Carlo simulation of $p\bar{p}$ events with $\sqrt{s}$ = 13 TeV in the ATLAS detector. These methods augment the already rich suite of $b$-tagging algorithms adopted by the ATLAS collaboration, by extracting complementary features and enhancing the overall flavor tagging performance.

ATLAS analysts can also expect an increased level of agreement between the train (Monte Carlo) and test (data) domains due to improvements in the tracking simulation with respect to the 2016 configuration, which reduces the performance losses associated with out-of-domain model application. 

\section*{Acknowledgements}
This work was supported in part by the Office of High Energy Physics of the U.S. Department of Energy under contract DE-FG02-92ER40704.


\section*{References}
\bibliography{main}

\end{document}